\documentstyle[12pt,psfig]{article}
\begin{document}

\title{Transparent Lyotropic Ferronematics}
\author{
V. V. Berejnov$^*$ and Yu. L. Raikher$^{**}$ \\
$^*$~\em Department of Chemical Engineering,\\[-8pt]
\em Technion -- Israel Institute of Technology, Haifa 32000, Israel \\
$^{**}$~\em Institute of Continuous Media Mechanics, \\[-8pt]
\em Ural Division of the Russian Academy of Sciences, Perm 614013, Russia}

\maketitle

\begin{abstract}
Transparent lyotropic ferronematic dispersions are synthesized by admixing
a cationic ferrofluid to the lyotropic liquid crystal PL/1D/Wt. The transparency
is spontaneous and time-independent. It is observed in zero magnetic field in
white and polarized light for disordered layers of ferroliquid crystals of
thicknesses $\geq 1$\,cm in both isotropic and nematic phases. Justification
of the low extinction of light in the systems of the lyotropic origin is
presented.
\end{abstract}

\newpage
\section*{Introduction}
The goal of this letter is to discuss the nature of the transparency effect
in lyotropic ferronematics. In general, the term {\em ferronematic} refers to
a magnetically-sensitive complex fluid resulting from mixing together magnetic
and liquid-crystalline components \cite{BrochDeGennes}. At present, three kinds
of those media can be distinguished. Namely, ferronematics can be synthesized
by admixing of ({\em i\/}) a paramagnetic ion dopant to a thermotropic liquid
crystal \cite{Ying2000}, ({\em ii\/}) a ferrocolloid to a thermotropic liquid
crystal \cite{Koner96}, and ({\em iii\/}) a ferrocolloid to a lyotropic
liquid crystal \cite{BerHigh}. A ferrocolloid (often called {\em ferrofluid})
is a colloidal dispersion of superparamagnetic nanoparticles in a liquid
carrier. The problem of primary importance for ferro-liquid-crystalline
compositions is their ability to resist the spontaneous aggregation of the
components. Recently, when solving the problem of colloidal stability for
lyotropic ferrodispersions with high concentration of magnetic particles,
we observed stable ferronematics, which were completely transparent in
a white and polarized light \cite{BerHigh,BerCol,BerPhysChem}.

The phenomenon of the liquid crystal transparency in thick samples ($\sim
1$\,cm) is an issue of considerable interest \cite{YamTan2001}. It is well
known \cite{dGenProst} that high turbidity of thermotropic liquid crystals
is due to the director fluctuations so that non-oriented thermotropics
completely depolarize and disperse light even in thin ($\sim 0.1$\,mm)
layers. With lyotropics we observed liquid crystals, which, being in the
non-oriented state, completely depolarize light but are transparent.
It is well known that occurrence of a similar effect in colloid systems
means a low level of aggregation, i.e., the stability of the system.
This property of ferrolyotropic dispersions helps a lot when investigating
their phase diagrams \cite{BerCol,BerPhysChem}.

A ferrolyotropic dispersion inherits transparency from both its components:
a ferrofluid and a pure lyotropic solution PL/1D/Wt, immaterial in which of
the two states---nematic or isotropic---the latter is. This transparency is
preserved in a broad temperature and concentration ranges and is long living.
In below we present a qualitative discussion of the phenomenon by comparing
light scattering intensities of classical thermotropics with those for
lyotropic liquid crystals and colloidal magnetic dispersions.

\section*{Experimental}
We use a lyotropic liquid crystal PL/1D/Wt, which is a ternary solution of
potassium laurate (PL), 1-decanol (1D) and water (Wt). The said solution
makes a lyotropic base for preparation of ferronematics. Potassium laurate
is synthesized in the laboratory by alkalinization of lauric acid with
potassium hydroxide. The details of synthesis are described in Ref.\
\cite{BerPhysChem}. At $20$\,C the concentration ranges are normally
5--8.8\,wt\%{} for 1D and 62--68\,wt\%{} for water with the saturation pH
value $\sim10$. The obtained samples are completely transparent and uniform
in both isotropic and nematic phases. The liquid crystal orients
spontaneously, which process takes minutes in a thin ($\sim 0.1$\,mm) glass
capillary and a week in a thick ($\sim 10$\,mm) glass tube. In non-oriented
layers, these lyotropic solutions remain entirely transparent but depolarize
the light completely \cite{BerPhysChem}, see the Fig.\ \ref{fig:1}(a,b).

For the studied PL/1D/Wt compositions, the clearing points are not observed,
neither with respect to temperature nor to concentration. In the nematic
region at 20\,C, a positive uniaxial optical anisotropy is detected.
Positive birefringence implies that the mesophase is organized as a discotic
structure. The micelle concentration in the lyotropic solution is about
$10^{19}$\,cm$^{-3}$. For similar systems in the same temperature range,
the micelle diameter and thickness were found earlier by means of
small-angle neutron scattering to be $\sim6.4$\,nm and $\sim 2.3$\,nm,
respectively \cite{Hendrix}.

Cationic ferrofluids are synthesized by the method described in Ref.\
\cite{MasrtCab}. They are stable aqueous colloids of $\gamma$-Fe$_2$O$_3$
under standard conditions: $pH \sim2$ and temperature 20\,C. Saturation
magnetization of the dispersed maghemite particles is about $\sim 300\,G$
\cite{Gazeauetal}. Their distribution is fairly well described by
a log-normal function with the characteristic diameter 7--8\,nm and the
dispersion $\sim0.3-0.4$. Concentration of the magnetic phase in the parent
ferrofluid was about $\sim5$\,vol.\%{}. The ferrofluids were optically
isotropic under zero external magnetic field.

Lyotropic ferrodispersions are prepared with the method described in Ref.\
\cite{BerCol} that yields transparent and stable ferroliquid crystals, see
Fig.\ \ref{fig:1}\,(c,d). The content of the magnetic component was
10$^{-4}$--1\,vol.\%{}. Characteristic red-brown color due to the presence
of ferroparticles was observed for all the samples. Strong
magneto-orientational effects at 10--100\,Oe were detected for ferronematic
layers of the thickness 0.1-10\,mm, see Refs.\ \cite{BerHigh,BerLett}.

\section*{Light Scattering by Liquid-Crystalline Materials}
Let us begin with the case a thermotropic nematic liquid crystal. In the
isotropic phase, the main part is played by the density fluctuations in the
liquid. In the ordered phase, the principal contribution goes from the
director fluctuations. After Ref.\ \cite{dGenProst}, the ratio of scattering
cross-sections in the nematic and isotropic states is
\begin{equation}                                                \label{eq:01}
\sigma_{\rm nem}^{(t)}/\sigma_{\rm iso}^{(t)}\simeq
\left[\varepsilon_{a}^{(t)}/{\varepsilon'}^{(t)}\right]^2\;
\left[K^{(t)}q^2\kappa^{(t)}\right]^{-1};
\end{equation}
here $q\sim2\pi/\lambda$, where $\lambda$ is the wavelength. In Eq.\
(\ref{eq:01}) the dielectric anisotropy $\varepsilon_{a} ^{(t)}$ and the
reference elastic constant $K^{(t)}$ (Frank modulus) are the parameters
inherent to the ordered state while the isothermal compressibility $\kappa
^{(t)}$ and the parameter ${\varepsilon'}^{(t)}$ characterize the isotropic
state. Note that hereafter we mark the material parameters referring to
thermotropic/lyotropic substances by superscripts ``$(t)$'' or ``$(\ell)$''.
For the order of magnitude estimate one can take  $\varepsilon_{a}^{(t)}\sim
{\varepsilon'}^{(t)}$.

With the same accuracy, one may set $\kappa\sim a^3/U$, where $U$ is the
typical energy of the molecule coupling and $a$ is the reference molecular
length, i.e., the size of the nematogenic chemical unit. For the elastic
constant, in the same way one gets $K\sim U/a$. Given that, Eq.\
(\ref{eq:01}) reduces to
\begin{equation}                                                \label{eq:02}
\sigma_{\rm nem}^{(t)}/\sigma_{\rm iso}^{(t)}\simeq1/(qa)^2.
\end{equation}
For small scattering angles: $q<10^{-2}$\,nm$^{-1}$.

In thermotropics, the structural unit is a single molecule with
$a\sim$\,2--3\,nm. In result, from Eq.\ (\ref{eq:02}) one finds that the
ratio (\ref{eq:02}) is about $10^6$. This estimation, first given in Ref.\
\cite{dGenProst}, shows that the nematic phase of a thermotropic mesogenic
substance is six orders of magnitude more opaque than its isotropic phase.
This explains why thermotropic mesogens, on entering the ordered state,
become completely turbid.

In a lyotropic solution, the elementary scattering unit is a micelle that
floats in an isotropic liquid carrier. This entails a considerable
difference between the light scattering mechanisms in lyotropic and
thermotropic nematogens. Each micelle is an aggregate comprising a number
of tensioactive molecules. In the considered here case of PL/1D/Wt, the
micelles are anisometric with the maximum size $a$ about $\sim5$\,nm.
However, despite being a multi-molecular object, a micelle still remains
an entity, whose size is much smaller than the optical wavelength. This
enables one to treat a lyotropic nematogen as a continuous medium.

Assuming, as in the case of thermotropics, that in the ordered phase the
main scattering is due to the director fluctuations, we retain in the
cross-section ratio $\sigma_{\rm nem}^{(\ell)}$ only the term with the
dielectric permeability anisotropy. Meanwhile, in the isotropic phase
a lyotropic solution by its scattering abilities cannot differ much from the
isotropic phase of thermotropic nematogens. Indeed, this scattering is caused
by the long-wave density modulations, i.e., is determined by the volume
elasticity (compressibility) of the liquid. Thence, constructing in the
same way as in Eq.\ (\ref{eq:01}) the cross-section ratio for a lyotropic
case, one gets
\begin{equation}                                                \label{eq:03}
\sigma_{\rm nem}^{(\ell)}/\sigma_{\rm iso}^{(\ell)}\sim\left
[\varepsilon_{a}^{(\ell)}/{\varepsilon'}^{(\ell)}\right]^{2}\;
\left[K^{(\ell)}q^2\kappa^{(\ell)}\right]^{-1}.
\end{equation}

The most important difference between thermotropic and lyotropic systems is
that, experimentally, all the known lyotropics are low-birefringent: their
dielectric anisotropy $\varepsilon_{a}^{(\ell)}$ is about $10^{-3}$ of their
isotropic dielectric constant.

Relating the reference values of the scattering ratios for thermotropics
and lyotropics under assumption that ${\varepsilon'}^{(t)}\sim{\varepsilon'}
^{(\ell)}$, one gets
\begin{equation}                                                \label{eq:04}
\sigma_{\rm nem}^{(t)}/\sigma_{\rm nem}^{(\ell)}\sim
\left[\varepsilon_{a}^{(t)}/\varepsilon_{a}^{(\ell)}\right]^{2}\;
\left[K^{(\ell)}/K^{(t)}\right]\phi^{-1},
\end{equation}
where the factor $\phi$ that is the volume fraction of micelles in the
solution. It has appeared due to the fact that in the ordered state only
the scattering from micelles is taken into account; we remark that in the
PL/1D/Wt systems under study we have $\phi\sim0.1$. The ratio $\varepsilon
_{a}^{(t)}/\varepsilon_{a}^{(\ell)}$ may be expressed in terms of the
refraction indices. We set $\varepsilon_a^{(\alpha)} =[n_e^{(\alpha)}]^2-
[n_o^{(\alpha)}]^2\sim n_o^{(\alpha)}\Delta n^{(\alpha)}$, where index
$\alpha$ marks either of the nematics. For thermotropics one has $\Delta
n^{(t)}\sim10^{-1}$ and $K^{(t)}\sim10^{-7}$\,dyn whilst for lyotropics:
$\Delta n^{(\ell)}\sim10^{-3}$ and $K^{(\ell)}\sim 10^{-6}$\,dyn;
the values of $n_o$ for both substances are of the same order. This casts
the ratio (\ref{eq:04}) in the form $\sigma_{\rm nem}^{(t)}/\sigma_{\rm
nem}^{(\ell)}\sim10^5/\phi$. As is shown in above, for a thermotropic
nematogenic substance the scattering ratio is $\sim10^6$ by itself,
therefore for a lyotropic system the corresponding ratio, see Eq.\
(\ref{eq:03}), is $\sim10\phi\sim1$.

Thus we conclude that in the isotropic state thermotropic and lyotropic
nematogens are equally transparent. On entering the nematic phase,
a thermotropic nematogen becomes $10^6$ times more opaque whereas the
transparency of a lyotropic liquid crystal is almost unaffected by the
orientational phase transition. This falls in full agreement with the
experimentally evidenced fact that even at a considerable sample thickness
(tubes $\sim1$\,cm), lyotropic liquid crystals are completely transparent
both in isotropic and nematic states.

In the case of a lyotropic ferronematic, the contribution to the light
scattering that owes to the presence of magnetic grains should be taken into
account. The main governing parameters are: the mean particle size $d$,
particle number concentration $c$ and aggregation number $m$. The first
two may be measured directly, and for our systems they are $d\sim8$\,nm and
$C\sim10^{14}-10^{15}$\,cm$^{-3}$.

To estimate the aggregation number, it is necessary to select the appropriate
scenario of aggregation in the lyotropic ferronematic. In the studied case,
the magnetic dispersions are rather dilute: the volume fraction of particles
does not exceed 0.01\,vol.\%{} and so is far from any of the phase separation
thresholds, see the diagrams in Refs.\ \cite{BerCol,BerPhysChem}. Under such
conditions, the main cause of aggregation is the interparticle magnetic
dipole-dipole interaction \cite{CabuilCous}, the intensity of which is
measured by the value of the dimensionless parameter $\Lambda=\mu^2/k_BTd^3$
with $\mu$ being the particle magnetic moment and $T$ the temperature
\cite{Shlim}. For single-domain particles one has $\mu=(\pi/6)M_sd^3$,
where $M_s\sim300$\,G is the magnetization of the colloidal maghemite; at
room temperature $k_BT\sim4\cdot10^{-14}$\,erg.

According to Ref.\ \cite{PshenBuz}, depending on the value of $\Lambda$,
there are three qualitatively different modes of aggregation. At $\Lambda<1$,
the particles are well separated and do not aggregate. In the range around
$\Lambda\sim3$ a considerable number of quasi-spherical droplets comprising
several particles is present . At $\Lambda>10$, the particles form
multi-grain open loops and long chain structures. With the above-mentioned
mean particle diameter, for our systems estimation gives $\Lambda\sim3$ that
implies that the particles, in majority, are associated in quasi-spherical
droplets with the aggregation number $m\sim$\,4--6. Recalculating this into
an effective aggregate diameter, one finds $d^{(f)}\sim16$\,nm.

Let us compare the Rayleigh scattering parameter of a ferrofluid, $R^{(f)}$,
and that of a lyotropic colloid, $R^{(\ell)}$. The pertinent formula, see
Ref.\ \cite{Lyk} for details, simplified for the case of dilute colloids
reads
\begin{equation}                                                \label{eq:05}
R^{(f)}/R^{(\ell)}\simeq\left[ C^{(f)}/C^{(\ell)}\right]\:
\left[d^{(f)}/d^{(\ell)}\right]^6 (\delta n)^{-2},
\end{equation}
where $\delta n$ is the difference between the refraction indexes of the
micelle and the solvent, $C^{(f)}$ and $C^{(\ell)}$ are the number
concentrations of the scattering units. To estimate the ratio (\ref{eq:05}),
we take for the concentrations $C^{(f)}\sim10^{14}$\,cm$^{-3}$ and $C^{(\ell)}
\sim10^{19}$\,cm$^{-3}$, for the mean diameters $d^{(f)}\sim16\,$nm
(a magnetic particle aggregate) and $d^{(\ell)}\sim3\,$nm (a micelle),
and $\delta n\sim0.1$ for the refraction indices difference. Hence, for
the ratio (\ref{eq:05}) one finds $R^{(f)}/R^{(\ell)} \sim1$. This estimate
agrees well with the evidence presented in Fig.\ \ref{fig:1} and proves
explicitly that an admixture of magnetic nanoparticles up to
$\Phi\sim0.07\,vol.\%$ does not cause any significant changes in the
transparency of a lyotropic ferronematic in  comparison with that of
a ``clean'' lyotropic system.

\section*{Conclusion}
Synthesis of lyotropic nematic ferroliquid crystals, which are highly
transparent, is reported that is done by admixing a cationic maghemite
ferrofluid to the lyotropic solution PL/1D/Wt. According to qualitative
considerations, the transparency is ensured by two principal circumstances.
The first is quite universal and refers to the fact that light scattering is
proportional to the square of the optical anisotropy $\Delta n$. Since
$\Delta n$ is low in all lyotropic systems, accordingly, small is the
corresponding contribution to the scattering. The second factor is
synthesis-sensitive. If a ferrolyotropic dispersion is stable enough,
as it happens for our samples, the mean aggregation number of magnetic
nanoparticles does not exceed several units. Because of that, the light
scattering caused by the dispersed solid phase remains less or at the
same level as that of the micellar solution. Summarizing, one can
characterize the reported lyotropic ferronematics as anisotropic fluids,
which uniquely combine the low magnetic-field control over the optical
axis direction and the high optical transparency.

\section*{Figure Caption}
{\bf Fig.\ 1.} Specimens of lyotropic solutions: (a) pure lyotropic in
a white light, (b) pure lyotropic in a polarized light, (c) ferronematic
in a white light; a concentration of the magnetic phase $\Phi=0.04\,vol.\%$,
(d) ferronematic in a polarized light; the tube diameter
is about $1$\,cm.

\section*{Figures}

see the file.jpg in attach.

\begin{figure}[!h]
\label{fig:1}
\end{figure}
\end{document}